# Femtomolar-level detection of SARS-CoV-2 spike proteins using toroidal plasmonic metasensors

Arash Ahmadivand,[1][†][*] Burak Gerislioglu,[2][†] Zeinab Ramezani,[3] Ajeet Kaushik,[4] Pandiaraj Manickam,[5,6] and S. Amir Ghoreishi[7]

[1]*Department of Electrical and Computer Engineering, Rice University, 6100 Main St, Houston, Texas 77005, United States*

[2]*Department of Physics and Astronomy, Rice University, 6100 Main St, Houston, Texas 77005, United States*

[3]*Department of Electrical and Computer Engineering, Northeastern University, Boston, MA 02115, United States*

[4]*NanoBioTech Laboratory Department of Natural Sciences, Division of Sciences, Art, & Mathematics, Florida Polytechnic University, Lakeland, Florida 33805, United States*

[5]*Electrodics and Electrocatalysis Division, CSIR-Central Electrochemical Research Institute (CECRI), Karaikudi 630 003, Tamil Nadu, India*

[6]*Academy of Scientific & Innovative Research (AcSIR), Ghaziabad 201 002, Uttar Pradesh, India*

[7]*Faculty of Electrical & Computer Engineering, Science and Research Branch, Islamic Azad University of Tehran, Tehran, Iran*

*aahmadiv@rice.edu

†: Equal contribution

**Abstract:** Effective and efficient management of human betacoronavirus severe acute respiratory syndrome (SARS)-CoV-2 infection i.e., COVID-19 pandemic, required sensitive sensors with short sample-to-result durations for performing diagnostics. In this direction, one of appropriate alternative approach to detect SARS-CoV-2 at low level (fmol) is exploring plasmonic metasensor technology for COVID-19 diagnostics, which offers exquisite opportunities in advanced healthcare programs, and modern clinical diagnostics. The intrinsic merits of plasmonic metasensors stem from their capability to squeeze electromagnetic fields, simultaneously in frequency, time, and space. However, the detection of low-molecular weight biomolecules at low densities is a typical drawback of conventional metasensors that has recently been addressed using toroidal metasurface technology. This research reports fabrication of a miniaturized plasmonic immunosensor based on toroidal electrodynamics concept that can sustain robustly confined plasmonic modes with ultranarrow lineshapes in the terahertz (THz) frequencies. By exciting toroidal dipole mode using our quasi-infinite metasurface and a judiciously optimized protocol based on functionalized gold nanoparticles (NPs) conjugated with the specific monoclonal antibody of SARS-CoV-2 onto the metasurface, the resonance shifts for diverse concentrations of the spike protein is monitored. Possessing molecular weight around ~76 kDa allowed us to detect the presence of spike protein with significantly low LoD ~4.2 fmol.

Keywords: *SARS-CoV-2 spike protein; COVID-19 pandemic; terahertz plasmonic biosensors, toroidal metasurfaces, femtomole-level detection*.

Pioneering efforts in developing reliable and timely recognition of infectious virus using standard methods date back to the end of the previous century, when intermittent wide-spread epidemics from emerging viruses, such as HIV, severe acute respiratory syndrome (SARS) and Middle East respiratory syndrome coronaviruses (MERS), pandemic influenza H1N1, Ebola and Zika, now SARS-CoV-2 associated respiratory syndrome i.e., COVID-19,[1-3] occurred.[1-6] Studies have shown that most of these epidemics derived from the zoonotic animal-to-human transmission incidents.[7,8] In spite of continuous advancements in the development of novel diagnostic approaches in recent years, on every occasion, dramatic scarcities in rapid and accurate detection of infections have hindered the public health response to the evolving epidemiological threat. Considering COVID-19 pandemic,[1-3,7,8] there have been unceasing efforts to develop sensitive and selective recognition modalities using both standard and innovative approaches needed for timely diagnostics of COVID-19. CRISPR–Cas12-based detection,[3] quantitative real-time polymerase chain reaction (RT-PCR),[9,10] ELISA (the enzyme-linked immunosorbent assay),[11] point-of-care (POC) lateral flow immunoassay test,[12,13] CT-imaging,[14] and some hematology parameters are the primary focus for clinical diagnosis of the recent fatal virus. Although the accomplished progress in such a short duration is promising, most of these techniques are subject to common limitations.[15] Firstly, many of the available techniques have long turnaround times and they are very slow. Secondly, for example, the RT-PCR tests require certified facilities, sophisticated equipment equipped laboratories, and well-trained personal. Thirdly, in many cases, they suffer from the poor sensitivity where a detection of virus is recommended at picomolar (pM). Finally, testing of specific antibodies against SARS-CoV-2 in collected patient's or recovered individual's blood is an invasive approach for the assessment of the disease. Instead, saliva test has just been recommended as a decisive and non-invasive sensing procedure at the very early-stages of the virus,[16,17] which possesses immense potential to be utilized in optical sensors and analogous non-contact devices. Recently, several nanoscale integrated architectures based on optical and electronic systems have been introduced to detect SARS-CoV-2 spike protein with much higher sensitivity, such as photothermal biosensors using gold nanoparticles (NPs) aggregates[18,19] and gated graphene-enhanced field-effect transistor (FET)-based biosensor.[20] Generally, most of these devices have assisted reasonably sensitive detection of SARS-CoV-2 virus, in subsequent biological components, needed for COVID-19 diagnostics as a major component of pandemic management. However, there is a continuous demand for on-chip, label-free, selective, repeatable, cost-effective, ready-to-use, and ultrasensitive biosensors.

Focusing on photonics architectures, among diverse types of optical biosensors, plasmonic devices based on quasi-infinite metasurfaces are promising tools that have facilitated precise screening and recognition of diverse biomarkers through substantial field confinement at subwavelength geometries.[21,22] This intriguing feature allowed detection of targeted biomolecules in highly diluted solutions at low concentrations.[21-24] As a cornerstone of photonics, artificially engineered plasmonic metasurfaces have stimulated the development of efficient, label-free, non-contact, non-invasive, and non-poisonous, multi-analyte immunosensing tools, devised for rapid and real-time detection of the fingerprints of diverse infections and viruses,[25-27] which are particularly important for the diagnosis of diseases and routine point-of-care (POC) clinical evaluations. Their remarkable advantages aside, most of the presently available optical biosensing and immunosensing methods dramatically suffer from poor limit of detection (LoD) and moderate precision in the identification of ultralow-weight biomolecules at lower densities.[28]

Most of these drawbacks have recently been addressed using the concept of toroidal metasensor based on flatland plasmonic structures.[30-35] It is shown that these platforms are operating based on the third family of resonant modes, known as toroidal multipoles.[36,37] The observation and study of these resonant moments in static form was reported in the middle of the previous century,[38] which was primarily established based on the physics of elementary particles. Later, the toroidal electrodynamic framework was described based on classical electromagnetism,[39] and subsequently the excitation of strong dipolar modes was realized in three-dimensional (3D) artificial media across the microwave frequencies.[40,41] Driven by the ongoing race to miniaturization, researchers are now able to fabricate toroidal-resonant metadevices in exquisite architectures to function at much longer frequencies).[42-45] For example, in the THz regime, a common feature of toroidal metasensors is the ability of focusing the incident electromagnetic radiation within a tiny spot. Additionally, by considering the Eq. 2 in Ref. [29], these metamolecules support resonances that possess much higher sensitivity to the refractive index perturbations in the surrounding media. To capitalize on these proof-of-principle developments, newly, practical and accurate screening of envelope proteins of a specific virus (Zika virus envelope protein with the molecular weight of ~13 kDa) and antibiotic molecules (kanamycin sulfate or Kantrex, $C_{18}H_{36}N_4O_{11} \times H_2SO_4$, with the molecular weight of ~600 Da) have been reported based on this metasensor technology.[31,33-35] Though the detection of biomolecules with molecular weight less than <600 Da was testified based on bulky 3D metamaterials (*e.g.* hyperbolic metamaterials) at pM densities,[46] similar performance has never been experienced using conventional structures based on flatland meta-optics. The low molecular weight of these biomolecules distinctively clarifies the superior performance of these sensors compared to analogous planar metasystems in all range of frequencies. This also verifies that the proposed THz toroidal metasensors are qualified candidates for the detection of much heavier hormones and drugs, organisms, enzymes, and envelope proteins at very low concentrations.

In this Letter, to resolve the inherent drawbacks of COVID-19 diagnosis tools via detecting SARS-CoV-2 spike protein, we developed a THz plasmonic biosensor device based on toroidal dipole-resonant metamolecules that demonstrates extreme sensitivity to the presence of SARS-CoV-2 spike protein. Toroidal dipole-resonant metasurfaces exhibit unconventional spectral properties that feature low-radiative losses, low mode volumes, and ultranarrow spectral lineshapes through robust confinement of electromagnetic fields. Using these distinct advantages, we tailored a symmetric multipixel planar metamolecule to support a pronounced toroidal dipole at the sub-THz spectra (~0.4 THz). To improve the binding properties of the targeted biomolecules to the devised metasurface, we introduced functionalized colloidal gold NPs conjugated with the respective antibody and captured the spike proteins present in the sample. Since the proposed configuration is a quasi-infinite platform based on periodic arrays of resonant unit cells, the practical realization of a miniaturized, multiplexed, and on-chip immunobiosensing instrument is possible for a wide range of applications (*e.g.* POC). Ultimately, we evaluated the performance of the sensor device in highly diluted solutions and showed that the implemented tool provides femtomolar (fM)-level detection of SARS-CoV-2 spike proteins.

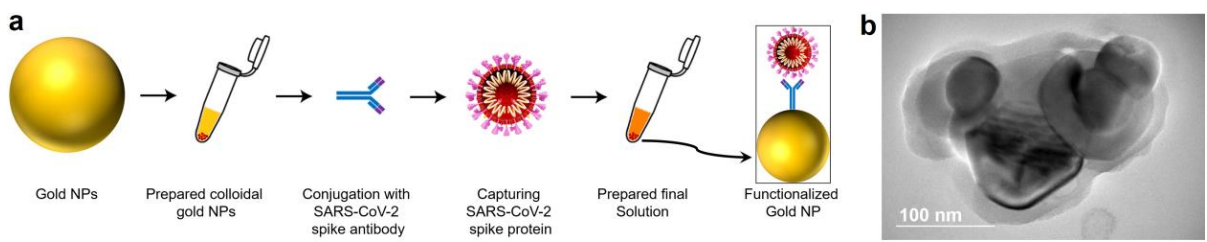

**Figure 1. Functionalization of colloidal gold NPs. a**, A schematic for the workflow of the developed functionalized gold NPs conjugated with the respective SARS-CoV-2 antibody and spike proteins. **b**, A TEM image of gold NPs conjugated with the respective antibody of SARS-CoV-2.

**Developed protocol to functionalize gold NPs.** A schematic of conducted research workflow is depicted in Figure 1a. Synthesized surface-functionalized gold NPs colloids, with the average diameter of ~45 nm (see Figure S1 in Supplementary Information), were utilized with robust covalent conjugation to primary amines (-NH2) of proteins. For the conjugation of SARS-CoV-2 (2019-nCoV, Rabbit MAb) Spike S1 antibody (Sino Biological Inc.) with the NHS activated gold NPs (1 mL), a reconstitution buffer was prepared by combining 0.1 M of reactant buffer with 50 µL of purified antibody. Right after the addition of buffer, the gold NP conjugate was and sonicated for 45 seconds. The solution is incubated while rotating at room temperature for 30 minutes. Next, 10 µL of quencher was added to deactivate any remaining active NHS-esters. The solution is centrifuged at 5700 rpm for 10 min (at 4 °C). Then, the supernatant is removed cautiously and resuspended with 0.1 mL of reaction buffer. By reiterating the centrifuging process with the same protocol, eventually, conjugate diluent (50 μL) was added and sonicated. The transmission electron microscopy (TEM) image of the functionalized gold NPs bounded with the respective antibody is depicted in Figure 1b. To prepare the gold NPs with specific antibodies [SARS-CoV-2 (2019-nCoV) Spike S1-His recombinant protein (Sino Biological Inc.)], we used both lyophilized 99% bovine serum albumin (BSA) (Sigma-Aldrich) and pH~7.4 phosphate buffer solution (PBS) to dissolve the immunoreagents. For real-time characterization, after washing the chips with PBS, antibody-modified structures were incubated in PBS containing 0.1 wt. % BSA for 30 min. Next, different concentrations of SARS-CoV-2 recombinant spike protein ranging from 2 fmol to 50 fmol in PBS was prepared using serial dilution. Once ready, the antibody-functionalized microstructures were rinsed and stored at 4 °C. An optimized time of 20 mins was used for the immucomplex formation at the miniaturized toroidal sensor metasurface. Figure 1a also contains an artistic picture of the functionalized gold NPs conjugated with the antibody and its subsequent binding with spike protein of SARS-CoV-2.

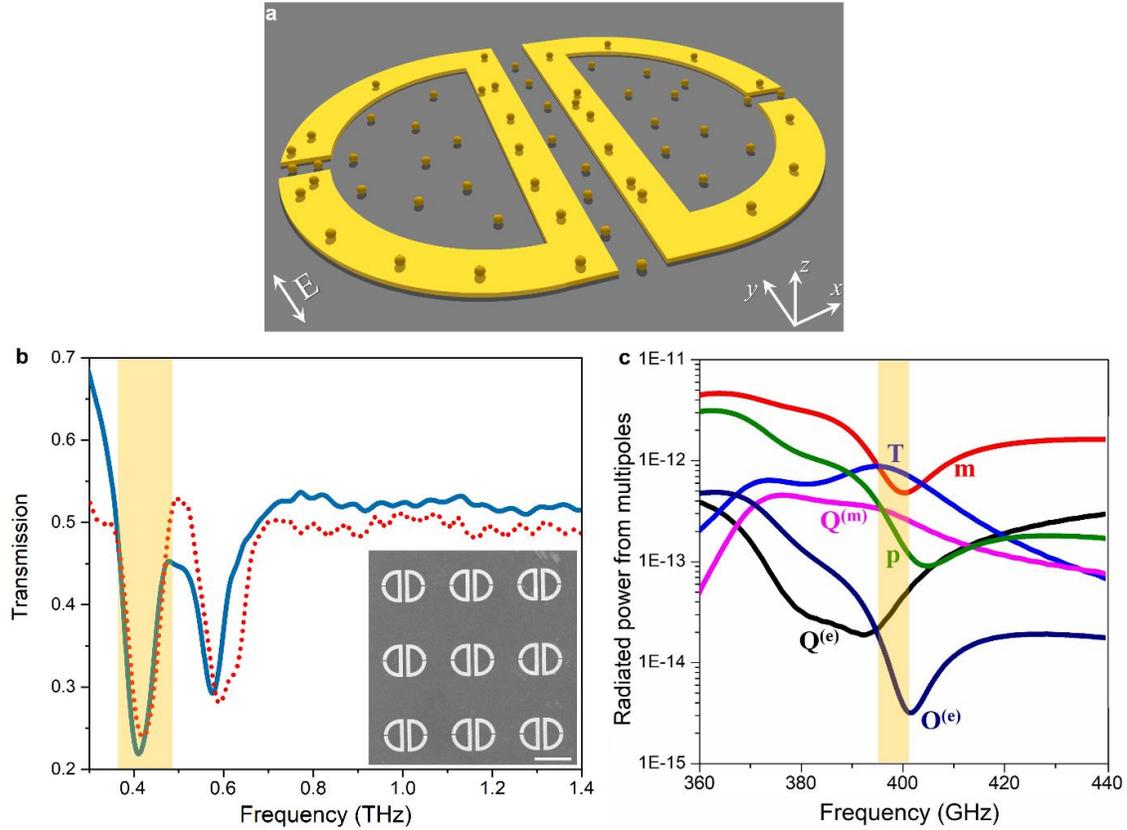

**Figure 2. Simulation, fabrication, and characterization of toroidal THz plasmonic metasurface. a**, An artistic rendering of the designed multipixel toroidal unit cell, including the dispersed functionalized gold NPs on top. The polarization direction of the incidence is demonstrated inside the panel. **b**, Numerically calculated (solid) and experimentally probed (dot) transmission spectra of the toroidal metasurface under transverse polarized THz light. The inset is the SEM image of the fabricated metasurface. Scale bar: 100 µm. **e**, Theoretically calculated scattered power from individual electromagnetic multipoles induced in the spectral response of metasurface. **p**: electric dipole, **m**: magnetic dipole, **Q$^{(e)}$**: electric quadrupole, **Q$^{(m)}$**: magnetic quadrupole, **T**: toroidal dipole.

**Spectral response of THz metasurface.** Figure 2a shows the schematic representation of the deigned toroidal THz metamolecule (not to scale), adopted from our previous work in Ref. [33], where functionalized gold NPs are dispersed onto the surface of the platform to improve the binding of biomolecules to the metasurface. The geometries of the unit cell are specified in the Supplementary Information (Figure S2). The numerically and experimentally studied transmission spectra (normalized) of the metasurface in the absence of gold NPs are plotted in Figure 2b. The spectra were obtained through the judicious geometric tuning of the unit cell under transverse-electric (TE) polarized illumination (*s*-polarized beam) to support a distinct toroidal dipole at the sub-THz spectra, around ~0.4 THz. Besides, this profile contains a second minimum around ~0.6 THz correlating with the magnetic dipole mode, which is not related to the scope of this study. The scanning electron microscope (SEM) image of the fabricated planar metasurface is shown in the inset of Figure 2b, which comprises periodic arrays of engineered gold metamolecules (details on numerical and experimental steps are explained in Supplementary Information). Considering the excitation principle of a toroidal dipole in planar metasurfaces,[47-49] we theoretically verified the formation of this mode via vectorial surface current density map to exhibit the required discrepancy between the direction of induced magnetic moments in proximal resonators. In particular, a destructive interference between these oppositely pointed magnetic moments in adjacent resonators leads to the formation of a head-to-tail configuration consisting of a confined arrangement of both charges and currents, which results in the projection of a dynamic toroidal dipole from the scatterer (Figure S3, Supplementary Information). To provide a full understanding of the induction of toroidal dipole resonance, we further analyzed the multipole decomposition of an arbitrary vector field by considering the radiated power from electric, magnetic, and toroidal multipoles (Figure 2c).[50] Noticing here, the toroidal dipole extreme is dominant contributor to the metasurface response, while the other dipolar and multipolar electromagnetic modes are suppressed.

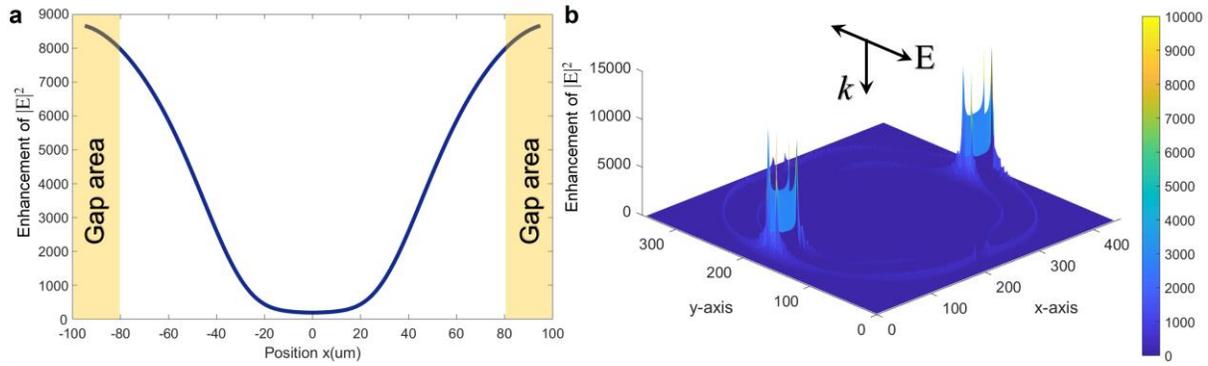

**Figure 3. E-field confinement in the toroidal metamolecule. a**, Cross-cut electric field enhancement ($|E|^2$) as a function of position along *x*-axis, simulated for a unit cell at the toroidal dipole frequency, showing significant intensity of the confined fields at the capacitive openings in each resonator. **b**, 3D E-field enhancement map at the frequency of toroidal dipole, in which the strong field confinement was spotted at capacitive gap regions. The polarization and incident light direction are indicated inside the panel.

Another key parameter of the resonant metasurfaces is the confinement of electromagnetic fields by the structured metamolecule. As previously discussed, the ultratight field confinement of toroidal metastructures is a unique feature that enables high sensitivity to the environmental perturbations, analogous to localized surface plasmons-based bio(chemical) sensors.[51,52] Figure 3a illustrates a cross-cut profile of electric-field (E-field) enhancement as a function of a unit cell length along the $x$-axis, showing a substantial field confinement at the capacitive gap regions in both resonators (see the highlighted regions in the panel). Beyond that, we calculated the 3D E-field enhancement map across the unit cell at the toroidal dipole frequency, validating the strong confinement of plasmons at the capacitive openings (Figure 3b). These locally generated energetic areas are related to the highlighted regions in the cross-sectional E-field enhancement map in Figure 3a.

**Detection of SARS-CoV-2 Spike Protein.** In continue, to highlight the important role toroidal radiations can play in the detection of targeted biomolecules, it is of value to define its behavior to the presence of SARS-CoV-2 spike proteins bounded to the metasurface, driven by an incident polarized beam. By putting our structured metasurface into practice through dispersing of the functionalized gold NPs onto its surface, we conducted a standard evaluation strategy to determine the analytic LoD and sensitivity of the device. To that end, we harnessed a protocol based on a combination of our previously utilized scenario for the diagnosis of Zika virus envelope proteins and screening of Kantrex biomolecules.[33,34] For the case of SARS-CoV-2 spike proteins, the assessment of these biomolecules surrounding the toroidal-resonant unit cell is executed by defining the difference between the transmission tensors in absence and presence of spike proteins. In this regime, we merely consider the tensors correlating with the transmitted and incident electric fields through the metasurface under $s$-polarized beam illumination ( $\Delta T(f) \equiv \left|T_{yy}^{\text{Antibody+PBS}}(f)\right|^2 - \left|T_{yy}^{\text{Spike protein}}(f)\right|^2$).[53] This model is built on the experimental evolution of the sensing approach. Using this mechanism, the transmission tensor due to SARS-CoV-2 spike proteins can be defined by subtracting following tensors: 1) the transmission tensors for the samples containing the respective antibody plus phosphate buffer solution (PBS), and 2) the same sample with the captured spike proteins. To this end, we initially determined the reference point for the proposed sensing technique by recording the transmission spectrum of the device by injecting PBS to the functionalized gold NPs conjugated with the relevant antibody. Later, we injected different concentrations of SARS-CoV-2 spike protein to the solution and measured the variations in the transmittance.

As discussed above, as a control, we primarily prepared 15 µL solution including functionalized gold NPs conjugated with the SARS-CoV-2 antibody and PBS (without the spike protein), then probed the transmission spectra (Figure 4a, black line). Repeating the same scenario for the samples in the presence of spike proteins, one can characterize the variations in the position of the toroidal dipole. The transmission profile for different concentrations of spike proteins (from 4 fmol to 12 fmol) is presented in Figure 4a, where the gold NPs play the fundamental role of conjugation with antibody, capturing of proteins, and binding with the metamolecules. Here, we observed a traditional and slight red-shift in the position of dipolar mode owing to surface chemistry. On the other hand, by dispersing the spike protein onto the metasystem, we observed a significant red-shift in the toroidal dipole frequency (Figure 4a), where a frequency shift of 2.12 GHz was observed for 4 fmol of spike protein. Continuous increases in the concentration of SARS-CoV-2 spike protein leads to additional red-shifts in the position of the resonance, and for 12 fmol, the frequency shift was measured around 6.912 GHz. It is noteworthy to consider that before each introduction of a new concentration of biomolecules, PBS was injected into the samples to remove the unbound or weakly attached spike proteins. Moreover, sensor performance was monitored as a frequency shift by dispersing of gold NPs with different concentrations of spike proteins (from 14 fmol to 20 fmol) onto the sensor surface. Nonlinear variation in the frequency shift (Figure 4b), as a function of the protein density, was observed, which allows to quantitatively determine the corresponding analytic LoD for the sensor around ~4.2 fmol based on a 4 times the signal-to-noise ratio (SNR) criterion. Various densities of functionalized gold NPs, including captured spike proteins, ranging from 2 to 20 fmol, were dropped and dried on the surface of the devised metasurface in several steps. The demonstrated findings in Figure 4b indicates the connection between the toroidal dipole shift and the protein concentration, in which 7.52 GHz shift in the position of the toroidal dipole was observed for 20 fmol of spike.

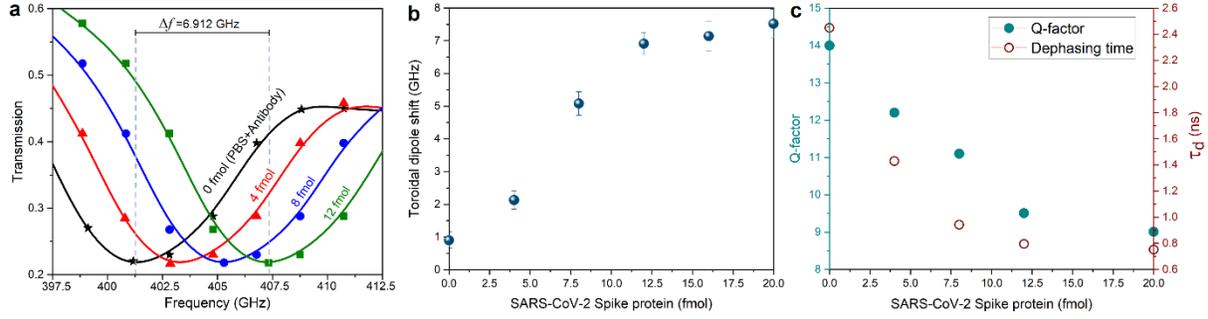

**Figure 4. Evaluation of metasensor performance using functionalized gold NPs and captured spike proteins. a**, Measured transmission spectra of the THz metasensor device for different concentrations (4 fmol to 12 fmol) of SARS-CoV-2 spike protein. For 0 fmol of spike protein, the spectral response was defined merely for antibody conjugated gold NPs plus PBS. **b**, Variations of the toroidal mode frequency shift for different concentrations of injected spike proteins captured by relevant antibody ranging from 2 fmol to 20 fmol. **c**, The $Q$-factor and dephasing time ($\tau_d$) of the toroidal lineshape as a function of protein concentration.

Such displacements in the position of toroidal dipole mode when the protein concentration increased accompanies with slight decline in its quality-factor ($Q$-factor). The variations in the $Q$-factor and corresponding dephasing time ($\tau_d$) due to the presence of spike proteins are demonstrated in Figure 4c. This panel illustrates the impact of the biomolecules in different condensation on the induced toroidal resonance lineshape and the corresponding dephasing time, because of the modified transmission tensor. Using the classical analysis for the calculation of radiative $Q$-factor, this component was defined as the ratio of the toroidal dipole frequency to the full width at the half maximum ($\delta$) of the lineshape ($Q=f_0/\delta$). On the other hand, following equation is used to quantify dephasing time:[54,55] $\tau_d= 2\hbar/\delta$. In Figure 5c, we plotted the radiative $Q$-factor and dephasing time variations as a function of SARS-CoV-2 spike protein concentration. Though by increasing the concentration of biomolecules, the linewidth of the toroidal dipole continuously decays, the induced resonance rationally retains its quality at high concentrations. This implies that at the higher densities of biomolecules, the developed metasensor keeps its performance.

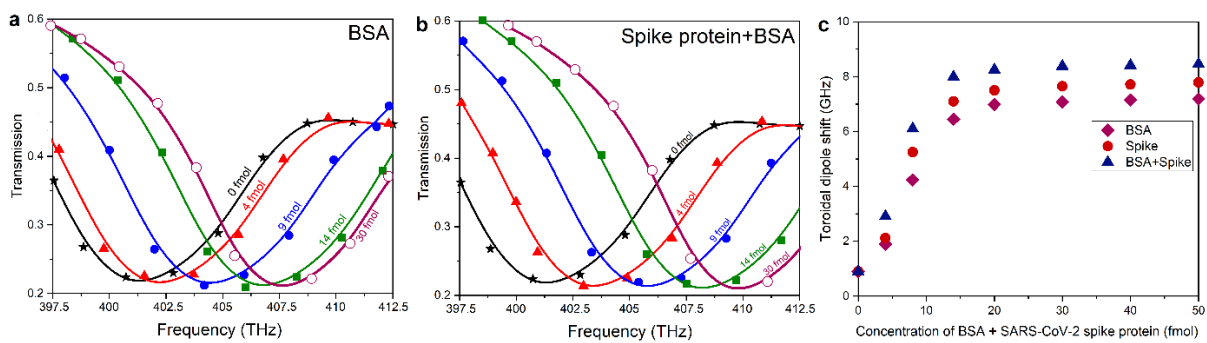

**Figure 5. Evaluation of the THz metasensor in diluted solution. a**, **b**, Measured transmission spectra of the THz metasensor device for different concentrations (from 4 fmol to 30 fmol) of BSA and SARS-CoV-2 spike protein + BSA, respectively. For 0 fmol of spike protein, the spectral response was defined merely for antibody conjugated gold NPs plus PBS. **c,** Variations of the toroidal dipole mode for different concentrations of BSA, SARS-CoV-2 spike protein, and BSA+SARS-CoV-2 spike protein. The spike proteins in all experiments were captured by functionalized gold NPs conjugated with the respective antibody.

Our findings by far raise the intriguing question: whether toroidal THz plasmonic immunobiosensor metasystem can be employed as a promising and precise tool to detect targeted biomolecules in highly diluted samples or not. To address this inquiry, and to qualitatively and qualitatively demonstrate the ultrahigh sensitivity of the implemented metasensor for the detection of spike proteins in extremely dilute solutions, we used a combination of biomolecules composed of high and low molecular weights. This includes targeted spike proteins plus BSA with molecular weights of ~76 kDa and ~66.5 kDa, respectively. Although the molecular weights of both biomolecules are close to each other, this allows to understand the influence of the heavy bio-objects on the sensing performance of the metadevice. Using the same protocol as we developed earlier in this study, we prepared colloidal gold NPs with different concentrations of BSA and spike protein. To carefully remove the unbounded and weakly conjugated/captured antibody and protein, as well as BSA molecules to the NPs, we thoroughly washed the samples before every experiment by distilled water. By introducing functionalized gold NPs solutions to the surface of the metastructure, we probed the transmission spectra as a function of BSA, spike protein, and spike protein + BSA at different concentrations (Figures 5a and 5b). Fixing the concentration of biomolecules in all assays between 2 fmol and 50 fmol, the acquired measurement results for the binding of BSA to colloidal gold NPs show the frequency shift owing to different concentrations of BSA. This is consistent with the observed nonlinear trend of the spike protein in Figure 4b. It should be underlined that direct measurement of the population of biomolecules on the metasurface is quite challenging, however, we can still define the impact of each biomolecule on the sensing performance quantitatively using the proposed methodology in Ref. [46]. By considering the shifts in the position of resonance for BSA biomolecules as 4.15 GHz and 6.19 GHz for spike protein at 10 fmol, this implies a mean shift per particle which can be approximated as 0.14 GHz and 0.21 GHz, respectively. It should be noted that as the molecular weight of spike protein is almost 1.15 times bigger than BSA, the relatively higher sensitivity going from BSA to SARS-CoV-2 spike protein is significant. Our experimental analyses also revealed that the sensor performance is high at the lowest concentrations, and as can be seen in Figure 5c, the shift in the position of toroidal dipole reduces dramatically and saturates for further concentrations of biomolecules. Ultimately, we summarized and listed the important parameters and results for the offered metasensor in Table 1. This allows to quantitatively compare the performance of the toroidal metasensor with recently reported similar mechanisms [15,17] for the detection of SARS-CoV-2 spike proteins.

**Table 1.** Important parameters and the obtained results from assays for the plasmonic metasensor.

| Sensor device | Plasmonic metasensor [Current approach] | Plasmonic photothermal biosensor [15] | Gated graphene-enhanced FET based biosensor [17] |
|---|---|---|---|
| **Target** | SARS-CoV-2 spike protein | SARS-CoV-2 spike protein | SARS-CoV-2 spike protein |
| **LoD** | ~4.2 fmol | 0.22 pM | 1 fg/mL |
| **Assay components** | Gold NPs, SARS CoV-2 antibody | Gold nanoparticles, SARS CoV-2 antibody | Graphene sheet, SARS CoV-2 antibody |
| **Preparation & Assay reaction time (estimated)** | ~50 min | > 3 h | N/A |
| **Assay sample-to-result time (estimated)** | ~80 min (including reaction time) | > 5 h | >48 h |
| **Assay Results** | Qualitative and Quantitative | Quantitative | Qualitative |

**Conclusion.** In conclusion, we have developed a highly sensitive miniaturized THz toroidal plasmonic metasensor to detect SARS-CoV-2 spike protein at femtomole-level concentrations. Our results revealed that introducing functionalized gold NPs to the multipixel metallic metasurface substantially improves the binding strength of biomolecules and subsequently boosts the sensitivity of device. Taking advantage of rationally high *Q*-factor, ultratight field confinement, and high sensitivity of toroidal resonances to the environmental perturbations enabled to devise a metasensor platform with unique LoD down to ~4.2 fmol. The performance of the plasmonic sensor device was carefully analyzed in highly diluted solutions containing BSA biomolecules and spike protein. This set of studies proved that the proposed metasensor can be utilized in the detection of SARS-CoV-2 spike proteins in real assays with fairly short sample-to-result duration (~80 min). We envision that the proposed non-invasive, non-contact, and rapid modality in this work allows for the diagnosis of SARS-CoV-2 spike protein at very early-stages with high precision.

**Outlook.** Our developed on-chip and miniaturized SARS-CoV-2 immunosensor based on THz plasmonic toroidal metasurface is able to detect SARS-CoV-2 spike protein selectively at fmol densities. Such of sensitive system is required for the early-stage COVID-19 infection diagnostics, assessment of therapy efficacy and generating data with it clinical relevance to manage/understand epidemic. Beyond that, development of POC system to make COVID-19 diagnosis easy at every location will certainly be useful to manage the infection related diseases. The developed metasensor device exhibited excellent performance in comparison to reported similar investigations in literature as summarized in Table 1. In future, our aim is to detect SARS-CoV-2 spike protein at POC after optimizing all the device components i.e., sensing chip (presented in this manuscript), full integration of metachip with microfluidic channel (in progress), and ultimately interfacing of analytical unit with smart display system (in progress). Ultimately sensing COVID-19 using smartphone based operation. However, the studied and presented sensor device is not yet tested using real samples due to lack of administrative formalities. Serious efforts are being made to establish a collaboration with U.S. based and overseas hospitals to arrange bio-fluids of COVID-19 infected patient. The results of the complete future studies related with POC sensing of COVID19 based on plasmonic and photonic metasensors will be published elsewhere.

## Acknowledgment

P. M. acknowledges the Start-up research grant (SRG/2019/000330) from Science & Engineering Research Board (SERB)-DST, Government of India.

## Authors' contributions

A.A. and B.G. equality contributed, and all other authors contributed extensively to the work presented in this article.

## Conflicts of interest

There are no conflicts of interest to declare.

## Data availability

All reagents used in this manuscript are available from commercial sources. In addition, the data that support the plots within this paper and other findings of this study are available from the corresponding author upon reasonable request.

## Code availability

The written and employed custom codes in FDTD program utilized in this study are available from the corresponding author upon reasonable request.